# Observations of wavefunction collapse and the retrospective application of the Born rule

Sivapalan Chelvaniththilan
Department of Physics, University of Jaffna, Sri Lanka
niththilan@univ.jfn.ac.lk

**Abstract**

In this paper I present a thought experiment that gives different results depending on whether or not the wavefunction collapses. Since the wavefunction does not obey the Schrodinger equation during the collapse, conservation laws are violated. This is the reason why the results are different – quantities that are conserved if the wavefunction does not collapse might change if it does. I also show that using the Born Rule to derive probabilities of states before a measurement given the state after it (rather than the other way round as it is usually used) leads to the conclusion that the memories that an observer has about making measurements of quantum systems have a significant probability of being false memories.

**1. Introduction**

The Born Rule [1] is used to determine the probabilities of different outcomes of a measurement of a quantum system in both the Everett and Copenhagen interpretations of Quantum Mechanics. Suppose that an observer is in the state $|M_B\rangle$ before making a measurement on a two-state particle. Let $|0\rangle$ and $|1\rangle$ be the two states of the particle in the basis in which the observer makes the measurement. Then if the particle is initially in the state

$$\frac{|0\rangle + |1\rangle}{\sqrt{2}}$$

the state of the combined system (observer + particle) can be written as

$$\left(\frac{|0\rangle + |1\rangle}{\sqrt{2}}\right)|M_B\rangle = \frac{|0\rangle|M_B\rangle + |1\rangle|M_B\rangle}{\sqrt{2}}$$

In the Copenhagen interpretation [2], the state of the system after the measurement will be either

$$|0\rangle|M_0\rangle \quad or \quad |1\rangle|M_1\rangle$$

with a 50% probability each, where $|M_0\rangle$ and $|M_1\rangle$ are the states of the observer in which he has obtained a measurement of 0 or 1 respectively. In the Everett interpretation [3], the final state is

$$\frac{|0\rangle|M_0\rangle + |1\rangle|M_1\rangle}{\sqrt{2}}$$

and the observer has a 50% probability of finding himself in either 'branch' of this state, i.e. in either $|0\rangle|M_0\rangle$ or in $|1\rangle|M_1\rangle$

Several authors have explored the reason for the Born rule, i.e. the reason why the square of the modulus of the amplitude is proportional to the probability. For example Deutsch [4] and Assis [5] show that it can be derived from game theoretical utility. Deutsch initially considers an equal amplitude superposition of two or more states of the observer and uses an essentially symmetry based argument to show that the observer must have equal probabilities of being in each branch.

Next, to deal with the case of unequal amplitudes he assumes that the amplitudes are in the ratio of the square roots of two integers, $\sqrt{m}:\sqrt{n}$. He then assumes that the first branch divided further into m branches and the second into n branches, thus reducing this problem to an equal amplitude one.

Pitowsky [6] also uses game theoretical methods to show that all aspects of quantum probability can be derived from rational probability assignments to finite quantum gambles.

Another approach, due to Zurek [7] uses the concept of envariance, i.e. environment assisted invariance. This is defined as a symmetry of the combination of a system and its surroundings in which the effect of a transformation acting solely on the system can be undone by another transformation acting solely on the surroundings. Zurek also first uses the properties of envariance to derive the Born rule for the case of equal amplitudes and then uses a method similar to that of Deutsch to generalise it to unequal amplitudes.

Carroll et al [8] follow a similar approach. They consider the time interval between the wavefunction branching via decoherence and the observer registering the outcome and show that the 'self-locating uncertainty' i.e. the probabilities for the observer to be in each branch are given correctly by the Born rule.

However Kent [9] criticises this method and also a similar method by Vaidman [10]. He points to one of the assumptions in the work of Carroll et al which is that in the time interval that they consider, there are several copies of the observer, one in each branch. He questions whether it is correct to call them different copies when they are in the same quantum state.

Rubin [11] uses the interpretation of probability as long term relative frequency. Building upon the work of Hartle [12], DeWitt [13] and Okhuwa [14], he considers an ensemble of identical systems and an observer making the same measurement on each system. In the limit of an infinite number of such systems, the amplitude of any Everett branch that does not obey the Born rule goes to zero.

These are just a few of the many works on the topic.

In this paper, I first consider the question of whether it is possible in principle to find out whether or not the wave function has collapsed. In Sections 2 and 3, I describe thought experiments that gives different results depending on whether or not a collapse has occurred. Then in Section 4, I explore what happens when the Born rule is applied retrospectively.

**2. A thought experiment with an electron**

When one observer makes measurements on a system, the probabilities of any outcome are the same irrespective of which interpretation is used, i.e. irrespective of whether or not the wavefunction collapses. But this is no longer true if there are two observers, one of whom can make measurements on the other. Deutsch [15] and Vedral [16,17] have made use of this fact to design thought experiments to find out whether a collapse has occurred.

Both their thought experiments involve one of the two observers making a measurement and then the second observer reversing the time evolution of the first observer, causing the measurement to be 'undone'. If the wavefunction of the first observer did not collapse as a result of the measurement, he is certain to end up back in the initial state when the measurement is undone, whereas if it did, he might also end up in a different state.

There are two difficulties with experiments of this kind. One is that the first observer must be kept completely isolated such that not even an atomic level interaction could happen with the environment. This is impractical as atoms of the observer will always be interacting via photons with those of the environment.

The second difficulty is in reversing the time-evolution that caused the measurement. If the Hamiltonian of the first observer is H during the measurement, which took a duration t to complete, then to reverse this process, the Hamiltonian must be made equal to -H for another time interval t. But the Hamiltonian contains kinetic energy terms of the form $p^2/2m$ and reversing this would mean introducing terms of the form $-p^2/2m$ which do not occur in any ordinary system.

I now show that it is possible to design a thought experiment that does not have this second difficulty, i.e. the collapse of the wavefunction can be detected without reversing the time-evolution of an observer. However, the first difficulty (isolation from the environment) still remains.

This thought experiment is based on the following fact about the Everett interpretation:

> *The average of the expectation values of a quantity in each branch <u>is not necessarily equal to</u> the expectation value of the same quantity in the superposition of all the branches.*

To see what this means, consider one observer, Bob who is totally isolated from the environment. Another observer Alice sends him an electron on which he performs a measurement. Alice can then make measurements on Bob or on the combined system of Bob and the electron.

We assume that the measurement device that Bob uses is designed to detect the x-spins of the electron and then to do nothing if the result is -1/2. But if the result is +1/2, it would then flip the spin of the electron to -1/2. The reason for designing the thought experiment in this way with the flipping process is that, as proved in a paper by Wigner [18], if we design it without the flipping process, an accurate measurement would only be possible if Bob is in a superposition of an infinite number of angular momentum eigenstates and this would complicate our calculations.

Let the electron be in the $|+z\rangle$ state initially and let $|M_B\rangle$ be the state of Bob and his measuring equipment before the measurement. Then the initial state of the combined system is

$$|+z\rangle|M_B\rangle = \left(\frac{|+x\rangle + |-x\rangle}{\sqrt{2}}\right)|M_B\rangle = \left(\frac{|+x\rangle|M_B\rangle + |-x\rangle|M_B\rangle}{\sqrt{2}}\right)$$

Let $|M_-\rangle$ and $|M_+\rangle$ respectively be Bob's states after he measures -1/2 or +1/2. Then in the Everett interpretation, the state of the combined system after the measurement and the conditional spin-flipping is

$$\left(\frac{|-x\rangle|M_+\rangle + |-x\rangle|M_-\rangle}{\sqrt{2}}\right) = |-x\rangle\left(\frac{|M_+\rangle + |M_-\rangle}{\sqrt{2}}\right)$$

since the process is designed such that the electron always ends up in the $|-x\rangle$ state. So the time evolution during the process can be expressed as

$$|+z\rangle|M_B\rangle \implies |-x\rangle\left(\frac{|M_+\rangle + |M_-\rangle}{\sqrt{2}}\right)$$

i.e.

$$\left(\frac{|+x\rangle + |-x\rangle}{\sqrt{2}}\right)|M_B\rangle \implies |-x\rangle\left(\frac{|M_+\rangle + |M_-\rangle}{\sqrt{2}}\right)$$

If instead the electron is in the $|-z\rangle$ state then by similar reasoning the time evolution is

$$|-z\rangle|M_B\rangle \implies |-x\rangle\left(\frac{|M_+\rangle - |M_-\rangle}{\sqrt{2}}\right)$$

i.e.

$$\left(\frac{|+x\rangle - |-x\rangle}{\sqrt{2}}\right)|M_B\rangle \implies |-x\rangle\left(\frac{|M_+\rangle - |M_-\rangle}{\sqrt{2}}\right)$$

All these are in the Everett interpretation. In the Copenhagen interpretation, the evolution is

$$|+z\rangle|M_B\rangle \implies |-x\rangle|M_+\rangle \quad or \quad |-x\rangle|M_-\rangle$$

with a 50% probability for each outcome and

$$|-z\rangle|M_B\rangle \implies |-x\rangle|M_+\rangle \quad or \quad |-x\rangle|M_-\rangle$$

again with a 50% probability for each outcome.

It is also easy to deduce the time evolutions for the cases where the electron is initially in the $|+x\rangle$ or $|-x\rangle$ states.

$$|+x\rangle|M_B\rangle \implies |-x\rangle|M_+\rangle$$

$$|-x\rangle|M_B\rangle \implies |-x\rangle|M_-\rangle$$

In these cases, the time evolution is the same in both interpretations.

Now let us apply the conservation of the z component of angular momentum to each of the above cases. Let the function $\langle J_z\rangle\{|\psi\rangle\}$ denote the expectation value of the z component of angular momentum for the state $|\psi\rangle$. Applying it to the $|+x\rangle$ case above gives

$$|+x\rangle|M_B\rangle \implies |-x\rangle|M_+\rangle$$

$$\therefore \quad \langle J_z \rangle \left\{ |+x\rangle \right\} + \langle J_z \rangle \left\{ |M_B\rangle \right\} = \langle J_z \rangle \left\{ |-x\rangle \right\} + \langle J_z \rangle \left\{ |M_+\rangle \right\}$$

Since both the the $|+x\rangle$ and $|-x\rangle$ states of the electron have zero expectation value of z-spin,

$$\langle J_z \rangle \left\{ |M_+\rangle \right\} = \langle J_z \rangle \left\{ |M_B\rangle \right\}$$

Similarly applying it to the $|-x\rangle$ case gives

$$\langle J_z \rangle \left\{ |M_-\rangle \right\} = \langle J_z \rangle \left\{ |M_B\rangle \right\}$$

Now let us apply the same to the $|+z\rangle$ case in the Everett interpretation.

$$|+z\rangle |M_B\rangle \implies |-x\rangle \left( \frac{|M_+\rangle + |M_-\rangle}{\sqrt{2}} \right)$$

$$\therefore \quad \langle J_z \rangle \left\{ |+z\rangle \right\} + \langle J_z \rangle \left\{ |M_B\rangle \right\} = \langle J_z \rangle \left\{ |-x\rangle \right\} + \langle J_z \rangle \left\{ \frac{|M_+\rangle + |M_-\rangle}{\sqrt{2}} \right\}$$

Since the $|+z\rangle$ state has an expectation value of +1/2 for the z-spin, this means

$$\langle J_z \rangle \left\{ \frac{|M_+\rangle + |M_-\rangle}{\sqrt{2}} \right\} = \langle J_z \rangle \left\{ |M_B\rangle \right\} + \frac{1}{2}$$

This means that in the Everett interpretation, if Alice measures the z component of Bob's angular momentum after he makes the measurement on the electron (which Alice sent him in the $|+z\rangle$ state) the expectation value of her measurement is

$$\langle J_z \rangle_{Everett} = \langle J_z \rangle \left\{ |M_B\rangle \right\} + \frac{1}{2}$$

Now to deduce what this expectation value will be in the Copenhagen interpretation, we cannot use the law of conservation of angular momentum. This is because this law is derived from the Schrodinger equation which, in the Copenhagen interpretation, is obeyed only between measurements but not during measurements (whereas it is always obeyed in the Everett interpretation). But we can deduce it by looking at the time evolution

$$|+z\rangle |M_B\rangle \implies |-x\rangle |M_+\rangle \;\; or \;\; |-x\rangle |M_-\rangle$$

with a probability of 50% each. Hence if the Copenhagen interpretation is correct, the expectation value of Alice's measurement on Bob is

$$\langle J_z \rangle_{Copenhagen} = \frac{1}{2} \times \langle J_z \rangle \left\{ |M_+\rangle \right\} + \frac{1}{2} \times \langle J_z \rangle \left\{ |M_-\rangle \right\}$$

$$\therefore \quad \langle J_z \rangle_{Copenhagen} = \langle J_z \rangle \left\{ |M_B\rangle \right\}$$

Hence this thought experiment gives different results depending on which 'interpretation' is correct. Hence strictly speaking, they are not two interpretations of the same theory but rather they are two different theories that make different experimentally falsifiable statements about the wavefunction, i.e. that it collapses or that it doesn't.

**3. A thought experiment with photons**

I now describe a similar thought experiment using photons rather than electrons. The advantage of this experiment is that if a large number of photons are used, the difference between the expectation values predicted by the Everett and Copenhagen interpretations will be comparatively large.

The following result will be needed for this thought experiment:

> *Any state of N photons with the same direction of propagation, in which the number of horizontally and vertically polarised photons is known with certainty, has an expectation value of zero for the total spin.*

In order to prove this, let us denote the horizontally and vertically polarised states of a single photon by $|H\rangle$ and $|V\rangle$ respectively. Let us also denote them by the following coloumn vectors.

$$|H\rangle = \begin{pmatrix} 1 \\ 0 \end{pmatrix} \quad and \quad |V\rangle = \begin{pmatrix} 0 \\ 1 \end{pmatrix}$$

Then the right and left handed circularly polarised photons respectively will be denoted by

$$|R\rangle = \frac{1}{\sqrt{2}} \begin{pmatrix} 1 \\ i \end{pmatrix} \quad and \quad |L\rangle = \frac{1}{\sqrt{2}} \begin{pmatrix} 1 \\ -i \end{pmatrix}$$

Consider the operator

$$S = \begin{pmatrix} 0 & -i \\ i & 0 \end{pmatrix}$$

Since

$$S|R\rangle = |R\rangle \quad and \quad S|L\rangle = -|L\rangle$$

and since the right and left handed circularly polarised photons have spins of 1 and -1 respectively, S is the spin operator. Note that

$$S|H\rangle = i|V\rangle \quad and \quad S|V\rangle = -i|H\rangle$$

Hence

$$\langle H|S|H\rangle = \langle V|S|V\rangle = 0$$

Hence both the linearly polarised states have zero expectation value of spin. The total spin operator for two photons is

$$S_2 = S \otimes \mathbb{I} + \mathbb{I} \otimes S$$

where $\mathbb{I}$ is the identity operator. Similarly the total spin operator for three photons is

$$S_3 \;=\; S \otimes \mathbb{I} \otimes \mathbb{I} \;+\; \mathbb{I} \otimes S \otimes \mathbb{I} \;+\; \mathbb{I} \otimes \mathbb{I} \otimes S$$

and so on. Let $|\psi\rangle$ be a state of N photons in which n are horizontally polarised (and N-n are vertically polarised). By looking at the forms of the operator $S_N$ in terms of S and $\mathbb{I}$ and noting that S transforms $|H\rangle$ and $|V\rangle$ into each other, we can deduce that $S_N|\psi\rangle$ must be a superposition of states that have either n-1 or n+1 horizontally polarised photons. Hence $S_N|\psi\rangle$ is orthogonal to $|\psi\rangle$ which means

$$\langle\psi|S_N|\psi\rangle = 0$$

Hence the expectation value of the total spin of $|\psi\rangle$ is zero and the required result is proved.

Now imagine that Bob is in an isolated box and the expectation value of his spin is $N_B$. Alice sends him N photons, all of which are right handed circularly polarised. Hence the total spin of these photons is N. Bob uses a polariser to measure the number n of photons that are horizontally polarised (and deduces that N-n are vertically polarised). Let us assume all the photons are absorbed during the measurement.

In the Everett interpretation the final state of Bob will be one in which he has a total spin of $N_B + N$ since he received N additional units of spin from the photons. But in the Copenhagen interpretation, his final state must have a total spin of only $N_B$ because when he measures the number of horizontally polarised photons, he collapses the state of the photons to one in which the expectation value of the total spin is zero.

**4.. The Born rule and false memories**

The Born rule is usually used to deduce the probabilities of different states after a measurement, given a particular state before the measurement. In this section, I will explore what happens if it is used the other way, that is, to deduce the probabilities of the states before the measurement, given a state after the measurement. We again assume that Bob measures the spin of an electron but we do not assume that his device conditionally flips its spin, since that is unnecessary for this calculation.

Let's say the electron is initially in the $|+x\rangle$ state and Bob first measures the x-spin and then the z-spin. In the first measurement, he is certain to get the result +1/2 while in the second measurement, he is equally likely to get +1/2 or -1/2. The time evolution of the system in the Everett interpretation during the two measurements can be represented as

$$|+x\rangle|M_B\rangle \;\;\Longrightarrow\;\; |+x\rangle|M_+\rangle \;\;\Longrightarrow\;\; \frac{|+z\rangle|M_{++}\rangle + |-z\rangle|M_{+-}\rangle}{\sqrt{2}}$$

where $|M_{++}\rangle$ is the state of Bob in which he remembers getting the result +1/2 for both measurements and $|M_{+-}\rangle$ is the state where he remembers getting +1/2 for the first measurement and -1/2 for the second. Since in the Everett interpretation, the time evolution is always unitary, there exist two unitary operators $U_1$ and $U_2$ that describe the evolution of Bob and the electron during the first and second measurements respectively. In terms of these operators, the time evolution can be written as

$$U_1 \,|+x\rangle|M_B\rangle = |+x\rangle|M_+\rangle$$

$$U_2 \,|+x\rangle|M_+\rangle = \frac{|+z\rangle|M_{++}\rangle + |-z\rangle|M_{+-}\rangle}{\sqrt{2}}$$

In the Copenhagen interpretation the evolution is

$$|+x\rangle|M_B\rangle \implies |+x\rangle|M_+\rangle \implies |+z\rangle|M_{++}\rangle \ or \ |-z\rangle|M_{+-}\rangle$$

Now consider the following hypothetical state:

$$|-x\rangle|M_+\rangle$$

This is an unusual state because even though the electron has a spin of -1/2, Bob clearly remembers having just measured its spin and obtained +1/2 as the result. It is therefore a state where Bob has a false memory. Leaving aside for the moment the question of how such a state can form, let us just note that such a state is nevertheless part of Bob's Hilbert space and hence it is meaningful to ask how it will evolve under the unitary operator $U_2$. The answer is easy to get if we express this state in terms of the z-spin of the electron:

$$|-x\rangle|M_+\rangle = \left(\frac{|+z\rangle - |-z\rangle}{\sqrt{2}}\right)|M_+\rangle$$

Since the process described by $U_2$ is the measurement by Bob of the electron's z-spin, we must have

$$U_2 \,|+z\rangle|M_+\rangle = |+z\rangle|M_{++}\rangle$$

$$U_2 \,|-z\rangle|M_+\rangle = |-z\rangle|M_{+-}\rangle$$

This implies that

$$U_2 \,|-x\rangle|M_+\rangle = \frac{|+z\rangle|M_{++}\rangle - |-z\rangle|M_{+-}\rangle}{\sqrt{2}}$$

The significance of such false memory states becomes clear when we think about Bob's measurement process in the Copenhagen interpretation:

$$|+x\rangle|M_B\rangle \implies |+x\rangle|M_+\rangle \implies |+z\rangle|M_{++}\rangle \ or \ |-z\rangle|M_{+-}\rangle$$

Consider the state $|+z\rangle|M_{++}\rangle$ above. This is one of the two possible states Bob can find himself in after the second measurement. Now assuming this is indeed Bob's state after the measurement, let us use the inverse unitary operator to find his possible states before it.

Since $U_2 \,|+z\rangle|M_+\rangle = |+z\rangle|M_{++}\rangle$, this means

$$U_2^{-1} \,|+z\rangle|M_{++}\rangle = |+z\rangle|M_+\rangle = \left(\frac{|+x\rangle + |-x\rangle}{\sqrt{2}}\right)|M_+\rangle$$

The Born rule now implies that if Bob's state after the second measurement is $|+z\rangle|M_{++}\rangle$, there is a 50% probability that his state before the second measurement was $|-x\rangle|M_+\rangle$, the false memory state. This line of thinking becomes more interesting if we go on to deduce what Bob's possible states before the first measurement. For this we will need the inverse of $U_1$.

Since $U_1 \,|+x\rangle|M_B\rangle = |+x\rangle|M_+\rangle$, this means

$$U_1^{-1}\,|+x\rangle|M_+\rangle = |+x\rangle|M_B\rangle$$

We also need to find $U_1^{-1}\,|-x\rangle|M_+\rangle$. Since $U_1$ describes a measurement of the x-spin of the electron, the x-spin eigenstates of the electron remain unchanged under this operator. Hence

$$U_1^{-1}\,|-x\rangle|M_+\rangle = |-x\rangle|M_B'\rangle$$

where $|M_B'\rangle$ is some other state of Bob. This is also a very unusual state because

$$U_1\,|-x\rangle|M_B'\rangle = |-x\rangle|M_+\rangle$$

which means that if Bob interacts with an electron in the $|-x\rangle$ state while he is in the $|M_B'\rangle$ state, he will end up with the false memory that he interacted with an electron in the $|+x\rangle$ state. Such states of Bob must exist in his Hilbert space.

Putting these together,

$$U_1^{-1}\left(\frac{|+x\rangle + |-x\rangle}{\sqrt{2}}\right)|M_+\rangle = \frac{|+x\rangle|M_B\rangle + |-x\rangle|M_B'\rangle}{\sqrt{2}}$$

which according to the Born rule means Bob has a 50% probability of having initially been in the $|M_B'\rangle$ state.

This problem of false memories is present not just in the Copenhagen interpretation. Consider the state after the second measurement in the Everett interpretation which is

$$\frac{|+z\rangle|M_{++}\rangle + |-z\rangle|M_{+-}\rangle}{\sqrt{2}}$$

This is an entangled state having two copies of Bob, one in the $|M_{++}\rangle$ state and the other in the $|M_{+-}\rangle$ state. Each of these copies has no way of knowing what state the other copy is in or the relative phase between the two copies. So for example, the copy of Bob in the $|M_{++}\rangle$ cannot know whether he is really in the above entangled state or perhaps in a different entangled state such as

$$\frac{|+z\rangle|M_{++}\rangle - |-z\rangle|M_{+-}\rangle}{\sqrt{2}}$$

If he is indeed in this entangled state, then his state before the second measurement must have been

$$U_2^{-1}\,\frac{|+z\rangle|M_{++}\rangle - |-z\rangle|M_{+-}\rangle}{\sqrt{2}} = |-x\rangle|M_+\rangle$$

which is the false memory state.

## 5. Discussion

The idea of false memory states seen above is closely related to the concept of Boltzmann brains. The Boltzmann brain paradox is a concept that arises when we try to account for the observation that the universe has far less entropy than we would expect from the principle of equal a-priori probabilities. One possible explanation, now believed to be the correct one [19] is that the universe began in a low-entropy state for an unknown reason and that its entropy has been increasing since then, but it has not reached thermal equilibrium.

Another theory, proposed earlier by Boltzmann [20] is that the ordered universe we observe is a result of an extremely rare thermal fluctuation from equilibrium. But Feynman [19] and Eddington [21] among others have pointed out that such a fluctuation that results in the entire universe becoming ordered is far less likely than a smaller fluctuation that results in the formation of just one conscious being (a Boltzmann brain) in a disordered, chaotic universe with false memories of having observed an ordered universe. The state $|-x\rangle|M_+\rangle$ discussed in the previous section is similar to a Boltzmann brain state.

Several solutions have been proposed for the Boltzmann brain paradox but it remains unsolved. For example, Boddy and Carroll [22] have explored whether vacuum instability induced by the Higgs field can prevent Boltzmann brains from forming. Carroll [23] has also speculated whether the idea of Boltzmann brains can be dismissed as being 'cognitively unstable.'

Many physicists including Carroll [23] have the view that Boltzmann brains can be used as a test of a cosmological theory: they argue that any theory in which Boltzmann brains greatly outnumber normal observers cannot be correct. Even though this seems reasonable, a mathematical proof of this statement has been difficult to provide.

Another view is that the cosmological measure must be chosen such that the probability of a typical observer being a Boltzmann brain is not too high. This assumption has been used by De Simone et al [24] and Linde et al [25-26] to study whether certain cosmological measures are acceptable.

Still it remains an unsolved problem.

## Acknowledgements

I am thankful to Dr William Matthews (Cambridge), Professor Oscar Dahlsten (SUSTech, China), Professor Vlatko Vedral (Oxford), Professor Davis Deutsch (Oxford) for valuable discussions.